\documentclass{article}

\usepackage[preprint]{nips_2018}

\usepackage[utf8]{inputenc} 
\usepackage[T1]{fontenc}    
\usepackage{hyperref}       
\usepackage{url}            
\usepackage{booktabs}       
\usepackage{amsfonts}       
\usepackage{nicefrac}       
\usepackage{microtype}      
\usepackage[pdftex]{graphicx}

\title{Prediction of Soil Moisture Content Based On  Satellite Data and  Sequence-to-Sequence Networks}

\author{
  Natalia Efremova\thanks{Centre for Corporate Reputation and Future of Marketing Initiative} \\
  University of Oxford\\
 \texttt{natalia.efremova2@sbs.ox.ac.uk} \\
  \And
  Dmitry Zausaev\\
  Deep Planet \\
   \texttt{dmitry@deepplanet.ai} \\
  \AND
  Gleb Antipov \\
  \texttt{gantipov@gmail.com} \\
}

\begin{document}

\maketitle

\begin{abstract}

The main objective of this study is to combine remote sensing and machine learning to detect soil moisture content. Growing population and food consumption has led to the need to improve agricultural yield and to reduce wastage of natural resources. In this paper, we propose a neural network architecture, based on recent work by the research community, that can make a strong social impact and aid United Nations’ Sustainable Development Goal of Zero Hunger. 
The main aims here are to: improve efficiency of water usage; reduce dependence on irrigation; increase overall crop yield; minimise risk of crop loss due to drought and extreme weather conditions. We achieve this by applying satellite imagery, crop segmentation, soil classification and NDVI and soil moisture prediction on satellite data, ground truth and climate data records. By applying machine learning to sensor data and ground data, farm management systems can evolve into a real time AI enabled platform that can provide actionable recommendations and decision support tools to the farmers.
 
\end{abstract}

\section{Introduction}
\label{intro}
The world's population is expected to rise from seven billion to ten billion in years to come. Fresh water scarcity and increased food consumption have lead to the need to increase agricultural yields up to 70 \% in the upcoming years [1].\\
Machine learning (ML) has emerged with big data technologies and high performance computing to create new opportunities in the agri-tech domain. ML tools can help to optimise the farming practice and enable sustainable use of environmental resources, e.g. without degrading the land but at the same time obtaining the most out of it.
Consequentially, this way of farming can also lead to restoration of environmental resources. \\
One way to help farmers is to provide specialised recommendations about their land, optimum type of crop, observed soil moisture, salinity, pH, and nitrogen content for fertilisation. If this information is  aggregated and analysed automatically at  a region-level, it can help to proactively take actions that can  increase agricultural yields, and, at the same time,  preserve scarce natural resources, such as fresh water, and minimise the usage of fertilisers and pesticides.\\
The advantage of using ML models versus manual labour is significant, since automated methods work in real time and cost significantly cheaper i.e. provide analysis of large territories much faster than conventional geographic information systems (GIS). Applying tools, which are flexible and adaptable, on large scale in precision agriculture gives an opportunity to have an integrated system on diverse levels of aggregation for the environmental resources. \\
In this work, we describe an ongoing project, that combines state-of-the-art neural network approaches to build an automated system for providing practical recommendations for farmers. In the rest of the paper, we  describe the model that we developed for soil moisture prediction on the farms and the requires data structure (section 2). We discuss the potential of the proposed system and outline the directions of future work in section 3. 

\section{Methods}
\label{methods}
\begin{figure}[ht]
    \begin{centering}
            {\includegraphics[width=0.8\textwidth]{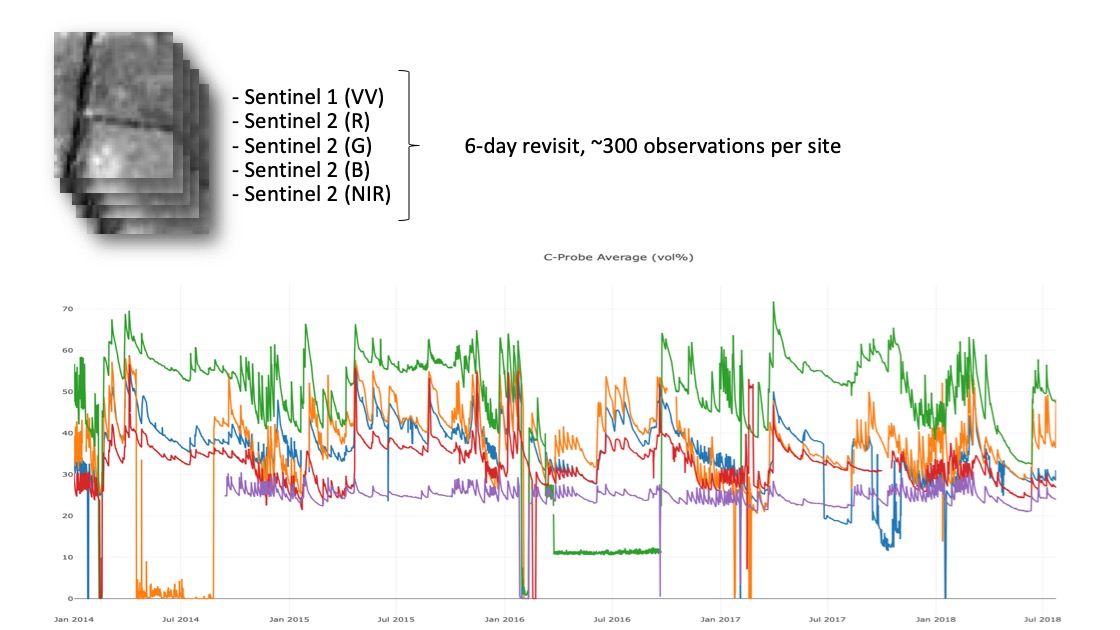}}
        \caption{Input data.We used 14 features from multiple channels (EO and ground data). We use dual (VV+VH or HH+HV) and single (HH or VV) polarisation from Sentinel-1, R,G,B and NIR bands from Sentinel-2 (top panel) and observations of soil moisture sensors from 20 sites from 3 wine regions in Australia; 7 years of daily data.}
        \label{fig:data_str}
    \end{centering}
\end{figure}

Soil moisture detection is one of the most important components of agricultural models, since it allows monitoring the state of the soil and water the crops. This paper offers a high-level architecture for vineyard management using free available satellite and in-situ sensing technologies. Specifically, the improvement of the efficiency in water use for irrigation to achieve a sustainable intensification of irrigated vineyards is nowadays a fundamental need. To plan the proper use of water and prevent over-irrigation and to generally control irrigated areas, Earth Obersevation (EO) data and in-situ sensing can be used to derive actual and forecasted crop water requirement maps, as well as soil moisture estimates.\\
We propose to use satellite imagery and radar data from the European Space Agency’s Sentinel-1 and Sentinel-2 satellite constellations for automatic feature extraction for soil moisture content (SMC) prediction. We compare the performance of 2 model, one of which includes ground measurements and the other one does not. \\
In water, molecular dipole moment oscillation, induced by falling electromagnetic radiation, produces polarization in reflected radiation. Therefore, satellite radar’s backscattering in polarized microwave bands depends on the SMC, incidence angle and landscape details (ground roughness and vegetation). Sentinel-1 is a constellation of two imaging Synthetic Aperture Radar (SAR) missions at C-band [2]. We use dual (VV+VH or HH+HV) and single (HH or VV) polarisation for SM modes. To increase accuracy and add automatic crop segmentation to the process, we use Sentinel-2 imagery (R,G,B, NIR bands) [3,4]. We propose automatic feature extraction based on the ground data observations and historical satellite imagery from Sentinel-1. We use additional satellite imagery and available ground measurements, such as weather station data and data from soil moisture sensors, to verify the accuracy of predictions. Usage of VV alone or the combination of VV and VH give similar accuracy on SMC estimates. We use NDVI values from Sentinel-2 imagery to increase the accuracy of our predictions (Fig.\ref{fig:data_str}). \\

\begin{figure}
    \begin{center}
    \begin{tabular}{cc}
        \includegraphics[width=0.4\linewidth]{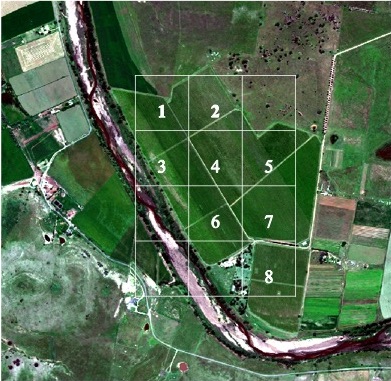}&
        \includegraphics[width=0.466\linewidth]{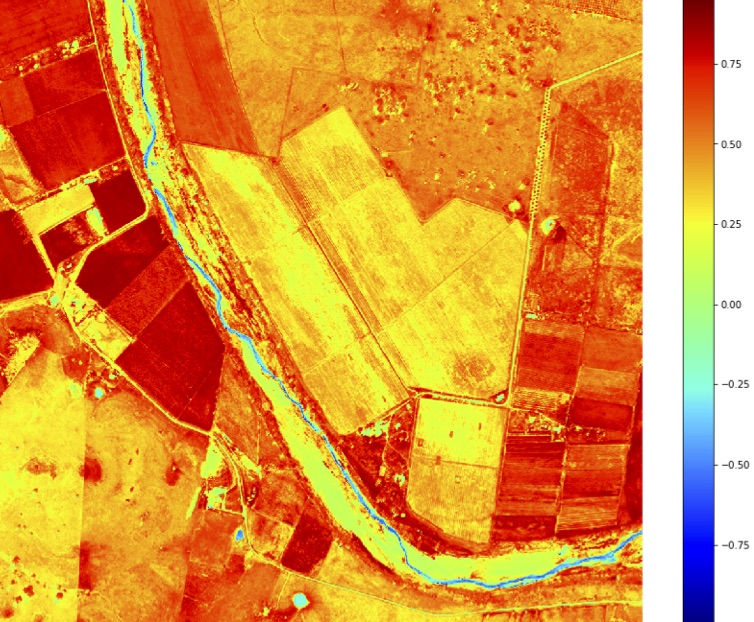}\\
        (a)&(b)\\
    \end{tabular}
        \caption{Test location for our model application (Hunter Valley, Australia): (a) location of the soil moisture sensors for each sort of the crop on the field: 1 - Glenesk Shardonnay R80, 2-Glenesk Shardonnnay R110, 3-Glenesk Shardonnay R132, 4 - Glenesk Shardonnay R192 and Mendoza, 5-Glenesk Merlot R165, 6-Glenesk Semillon R245, 7-Glenesk Shiraz R50, 8-Glenesk Verdelho R211; (b) heatmap for crop stress analysis (Normalized Difference Vegetation Index, NDVI).
 }
        \label{fig:mult}
        
    \end{center}
\end{figure}
\begin{figure}[ht]
    \begin{centering}
            {\includegraphics[width=0.8\textwidth]{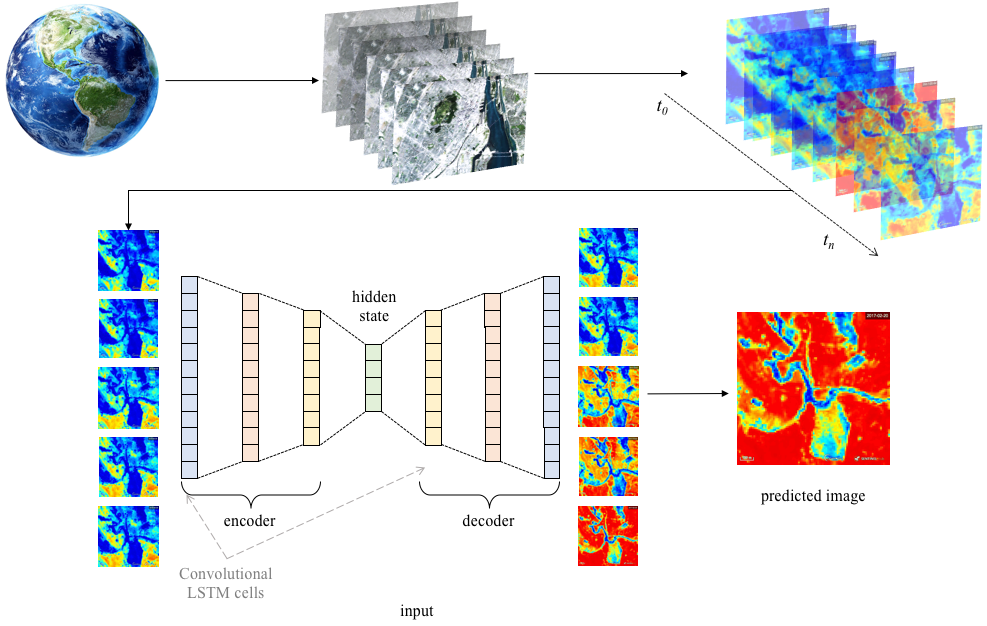}}
        \caption{ The architecture of a model for prediction of soil moisture content consists of two recurrent neural networks, combined in an encoder-decoder framework. The input image was flattened into the 1D array and processed with an encoder network. Convolutional LSTM cells are used in both encoder and decoder networks. The output of the model is the sequence of predicted images. Here, colour represents the amount of water in the soil from wet (blue) to dry (red).}
        \label{fig:autoenc}
    \end{centering}
\end{figure}

We use next image prediction with convolutional sequence-to-sequence autoencoder for prediction of soil moisture content from historical radar data (satellite aperture radar, SAR). Deep recurrent neural networks proved to be very useful in predicting sequences because they learn temporal dependencies in sequential data [5]. There are many examples of successful applications of seq2seq architectures to sequences of images [6], [7]. The type of sequence-to-sequence architecture, proposed in this work, was first applied to satellite imagery in [8]. To access satellite imagery, we used Sentinel Hub for bulk download of Sentinel-1 and Sentinel-2 data. As input features, we used a combination of bands (Sentinel 1 : C and X bands, Sentinel 2: combinations of R, G, B and near-infrared (NIR) bands). As a source of ground-truth data, we use soil moisture sensor readings and historical weather sensor data from a vineyard in Hunter Valley, Australia. The map of the vineyard we were working with with overlaid location of soil moisture sensors is depicted on Fig.\ref{fig:data_str}. We also use historical data, collected over 20 years, together with ground data from weather station and soil moisture sensors to provide ground-truth measurements and validation to our model. The proposed architecture consists of two recurrent neural networks, combined in an encoder-decoder framework (Fig.\ref{fig:autoenc}. The input image was flattened into the 1D array and processed with an encoder network. Convolutional long-short term (LSTM) cells are used in both encoder and decoder networks. The output of the model is the sequence of predicted images, which contain visual information about the amount of water in the soil. 

\begin{figure}[ht]
    \begin{centering}
            {\includegraphics[width=0.8\textwidth]{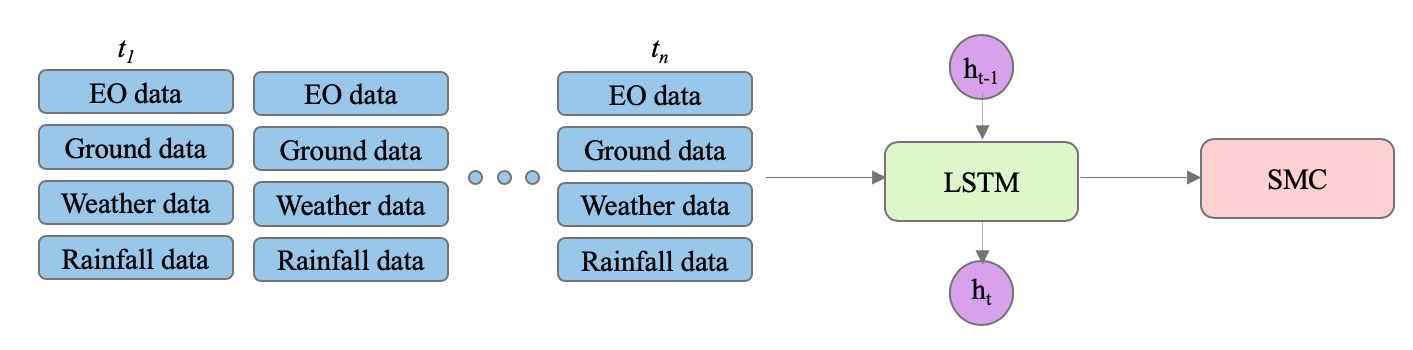}}
        \caption{The architecture of the LSTM model.The inputs to this model are radar reflectance data and ground SMC measurements. We use the data from two satellite constellations (Sentinel 1 and Sentinel 2) as input EO data and seven years of daily ground SMC measurements from fourteen sites for training. Additionally, we use publicly available weather data and rainfall data.}
        \label{fig:lstm}
    \end{centering}
\end{figure}

We compare two architectures for soil moisture content prediction AE and LSTM models. Our research suggests that we can obtain similar prediction accuracy with AE and LSTM architectures (Fig.\ref{fig:lstm}. LSTM architecture performs better with smaller amount of observations but requires soil moisture sensor (SMS) data, while AE works only with EO data, but requires significant amounts of historical data.

\section{Results and Discussion}
\label{discussion}
In this paper, we aim to demonstrate an application of AI that can help underpin the work needed to support the sustainability goal of zero hunger. Increasing agricultural yield and reduction of wastage of natural resources, such as water, can drive positive economical outcomes for the farmer e.g. reduced costs and investment on irrigation, soil moisture sensors and manual labour. On a macro scale, increasing agricultural yield and total factor productivity (TFP) of a country will assure that the needs of the growing population don't outstrip the ability to supply food. According to [10], growth in yield and labour productivity are highly associated with poverty reduction, but the extent to which they affect poverty sharply varies across regions.\\
Worldwide precision agriculture market is expected to reach approximately US\$ 7.9 billion by 2022 growing at a compound annual growth rate (CAGR) of 16\% [11]. While organisations and companies may be interested in the large financial value of this market US\$240B [11], there is also a growing trend of organisations working on global sustainability goals that not only aids social good on a large scale but also makes a positive impact on the firm's reputation. According to McKinsey [12], one top reason why organisations address sustainability is to "build, maintain or improve reputation". Therefore, application of AI to earth observation (EO) data has a significant potential on large scale as well on the small scale. On the large scale, it will benefit large organisation and country economy as a whole. At the same time, on the small scale it will help farmers to manage their land more efficiently and to gain more profits from their land.\\
The usage of earth observation (EO) data and ML-tools for analytics and forecasting thus significantly impacts the country and provides agricultural sustainability and productivity on global scale. Current method provides better recommendations for farmers. Such recommendations are particularly important in the regions where usage of measuring equipment for obtaining the information about soil or using drones for field monitoring is difficult due economic situation in the region. \\
We have tried our predictions on the medium-sized vineyard in Australia. Our experiments on the pilot farm have shown that the current state-of-the-art ML tools can effectively replace costly soil moisture sensors and are four times cheaper in implementation. It has been shown, that the usage of EO data and ML for analytics and forecasting can be valuable tools for agricultural yield prediction [13] and for poverty prediction [14].  We propose adding more to this functionality by soil moisture prediction, automatic crop detection and crop stress analysis.\\
The combination of the proposed tools provide visual aid and recommendations for farmers how to use their land in a the best way. We propose an intuitive interface that includes: detecting the current crops on the patch of land the farmer is located in; providing recommendations on actions, required for particular crops recommendations for using other crops for current state of the land (including soil analysis, water content analysis etc.), which will produce better yield while consuming less resources, scarce in particular region.


\section*{References}
\small
[1] Food and Agricultural Organisation of the United Nations (2009) How to Feed the World in 2050. \textit{Expert report}. 

[2] Snoeij, P. \ \& Attema, E. \ \& Torres, R. \ \&  Levrini, G. \ \& Croci, R. \ \& L'Abbate, M. \ \& Pietropaolo, A. \ \& Rostan, F. \ \& Huchler, M.\ (2009). Sentinel 1 - the future GMES C-band SAR mission.  

[3]	Paloscia, S.\ \& Pettinato, S.\ \& Santi, E.\ \& Notarnicola, C.\ \& Pasolli, L.\ \& Reppucci, A. \ (2013) Soil moisture mapping using Sentinel-1 images: Algorithm and preliminary validation. {\it Remote Sensing of Environment 134}, pp. 234 -- 248. 

[4]	Pasolli, L.\ \& Notarnicola, C.\ \& Bruzzone, L.\ \& Bertoldi, G.\ \& Della Chiesa, S.\ \& Hell, V.\ \& Vaglio Laurin, G.\ (2011) Estimation of Soil Moisture in an Alpine Catchment with RADARSAT2 Images. {\it Applied and Environmental Soil Science, 2011}, pp. 1 -- 12. 

[5] Sutskever, I. \ \&  Vinyals, O. \ \&  Le, Q. V.\ (2014) Sequence to sequence learning with neural networks. {\it Advances in Neural Information Processing Systems}, pp. 3104 -- 3112.

[6] Smilevsky, M.\ \&  Lalkovsky, L. \ \& Madjarov, G. (2018) Stories for images-in-sequences by using Visual and Narrative Components. ArXiv preprint: 1805.05622v3.

[7] Noroozi, M.  \ \&  Favaro, P. (2016) Unsupervised Learning of Visual Representations by Solving Jigsaw Puzzles. \textit{European conference on computer vision, ECCV 2016.}

[8] Hong,S. \ \&  Kim, S.\ \&  Joh, M.\ \&  Song,S-K. \ (2017) PSIque: Next Sequence Prediction of Satellite Images using a Convolutional Sequence-to-Sequence Network.  {\it Workshop on Deep Learning for Physical Sciences, NIPS 2017}.

[9] Ronneberger, O.\ \&  Fischer,P. \ \&   Brox, T. (2017) U-Net: Convolutional Networks for Biomedical Image Segmentation. \textit{MICCAI 2015: Medical Image Computing and Computer-Assisted Intervention} pp. 234 -- 241.

[10] Goldman Sachs (2016) Precision Agriculture: Cheating Malthus with Digital Agriculture. \textit{Profiles in Innovation.}

[11] Beige Market Intelligence (2017) Global Precision Agriculture Market - Strategic Assessment and Forecast 2017-2022. 141 p.

[12]  Bové, A. \ \& D'Herde, D. \ \& Steven S.\ (2017) Sustainability's deepening imprint. 
{\it McKinsey \ \&  Company, Insights on Sustainability \ \&  Resource Productivity}.

[13] You, J.\ \&   Li, X. \ \&   Low, M. \ \&   Lobell M. \ \&   Ermon S. (2017) Deep Gaussian Process for Crop Yield Prediction Based on Remote Sensing Data. \textit{Proceedings of the Thirty-First AAAI Conference on Artificial Intelligence (AAAI-17)}.

[14] Jean, N. \ \& Burke, M. \ \& Xie,M. \ \& Davis, W.M. \ \&   Lobell M. \ \&   Ermon S. (2016) Combining satellite imagery and machine learning to predict poverty
\textit{Science}.

\end{document}